\documentclass[10pt]{article}

\usepackage{graphicx,epsfig}
\begin{document}

\begin{titlepage}
\begin{center}

{\hbox to\hsize{
\hfill PUPT-1934}}
{\hbox to\hsize{
\hfill RU00-2-B}}

\bigskip

\vspace{4\baselineskip}

{\large \sc \bf Inflationary solutions in the brane-world and 
their geometrical interpretation \\}

\bigskip

\bigskip

\bigskip

\bigskip
\bigskip

{\sc  Justin Khoury$^1$, Paul J. Steinhardt$^1$ 
and Daniel Waldram$^{2,3}$}\\[0.5cm]

$^1${\it Joseph Henry Laboratories,\\
Princeton University,
Princeton, NJ 08544}\\[.4cm]

$^2${\it Theory Division, CERN CH-1211, Geneva 23, Switzerland}\\[.4cm]

$^3${\it Department of Physics, The Rockefeller University,\\ 
1230 York Avenue, New York, NY 10021}\\[1cm]

{\bf Abstract}\\
\end{center}
We consider the cosmology of a pair of domain walls bounding a
five-dimensional bulk space-time with negative cosmological
constant, in which the distance between the branes is not fixed in
time. Although there are strong arguments to suggest that this
distance should be stabilized in the present epoch, no such
constraints exist for the early universe and thus non-static solutions
might provide relevant inflationary scenarios. We find the general
solution for the standard ansatz where the bulk is foliated by
planar-symmetric hypersurfaces. We show that in all cases the bulk
geometry is that of anti-de Sitter (AdS$_5$). We then present a geometrical
interpretation for the solutions as embeddings of two de Sitter
(dS$_4$) surfaces in AdS$_5$, which provide a simple interpretation of
the physical properties of the solutions. A notable feature explained
in the analysis is that two-way communication between branes expanding
away from one another is possible for a finite amount of time, after
which communication can proceed in one direction only. The geometrical
picture also shows that our class of solutions (and related solutions
in the literature) are not completely general, contrary to some
claims. We then derive the most general solution for two walls in
AdS$_5$. This includes novel cosmologies where the brane tensions are
not constrained to have opposite signs. The construction naturally
generalizes to arbitrary FRW cosmologies on the branes. 

\end{titlepage}

\newpage

\section{Introduction}

There has been considerable interest over the last few years in space-time
geometries where gauge and matter degrees of freedom are confined to a
four-dimensional submanifold, while gravity is allowed to propagate in the
whole of space-time. The motivation for such geometries comes from
Ho\v rava-Witten theory, or M-theory, compactified on
$S_1/Z_2$~\cite{witten}, or models built from D-branes. For instance, in
Ho\v rava--Witten theory, after compactification on a Calabi-Yau
three-fold, the five-dimensional vacuum solution is found to consist
of two three-branes, each coinciding with an orbifold fixed plane
\cite{lukas1}. The brane tensions have opposite sign, and their
magnitude is related to the bulk cosmological constant. The existence
of a higher-dimensional bulk is likely to have deep implications for
the dynamics of the early universe, in particular for inflation. 

More recently, Randall and Sundrum \cite{randall} proposed that the setup
of two domain walls bounding an orbifold could solve the hierarchy
problem. Cosmological solutions for similar geometries were presented in
Refs. \cite{kaloper, nihei, kanti}, but all assuming some potential that
stabilizes the orbifold separation. While a stabilizing potential
might be necessary to obtain a realistic picture of our universe
today, there is no reason to believe that the extra dimension was
static during inflation. In Ref. \cite{lukas}, general 
inflationary solutions were given for compactified Ho\v rava-Witten
theory, where the brane energy density is constant and the 
bulk cosmological constant is set equal to zero. For the case of negative
bulk cosmological constant, a class of solutions was derived in Ref.
\cite{kim} where the bulk geometry is pure anti-de Sitter (AdS) space.
However, the most general solution was not obtained.

In section 2, we find a class of solutions in the case
where the bulk geometry is AdS$_5$, the stress energy on the walls is
given by their tensions, and the walls are located at fixed orbifold
coordinates. The coordinate system used to find the solutions is not
the most general, but has the nice property of making homogeneity and
isotropy along the three spatial directions (i.e., planar symmetry)
manifest at each point along the orbifold direction and, in
particular, on each brane.  
While our ansatz coincides with that of Ref. \cite{kim}, our class of
solutions (described algebraically in section \ref{soln}) is obtained
by explicitly solving Einstein's equations, and hence is the most
general within the set-up described above. It is parameterized by a
few constants and an arbitrary periodic 
function of one variable, and is very similar in form to that found in
Ref. \cite{lukas} for the case of Minkowskian bulk geometry.

Since we are interested in inflationary solutions, we restrict to the
case where the tensions of the branes are greater in magnitude than
the bulk cosmological constant. With this minor assumption, our
solutions all describe the same cosmology on the boundaries, namely
that of a de Sitter (dS) phase. Thus, in order for the branes to
generate AdS in the bulk, it is necessary that they follow de Sitter
trajectories in AdS. It is therefore natural to interpret our
solutions as embeddings of two dS$_4$ surfaces in AdS$_5$, and this is
the focus of the second half of the paper. As a warm-up, in section
\ref{dynamics} we will describe how the solution of
Ref. \cite{lukas} (with zero bulk cosmological constant) describes slices of ${\cal M}_5$ (Minkowski
space). The generalization to AdS$_5$ is discussed in section
\ref{l<0}. A nice feature of this analysis is that it provides a
geometrical interpretation of the various arbitrary parameters
describing our class of solutions. It is also a useful tool to
describe the time-evolution of the orbifold as well as the causal
properties of the solutions (sect. \ref{dynamics} and \ref{l<0}). We
find that our solution describes scenarios where the orbifold
direction is static, collapses or expands, as seen from an observer
living on one of the branes. In solutions with expanding extra
dimension, two-way communication between the walls is only possible
for a finite amount of time. Afterwards, communication can proceed in
one way only: signals emitted from the negative-tension brane do not
reach the positive-tension brane in finite affine parameter. 
 
Finally, the geometrical picture allows us to realize that the
embeddings of section 2 display an unexpected amount of symmetry,
leading us to postulate that our solutions are in fact a subclass of
all possible embeddings of two dS$_4$ in AdS$_5$. In section
\ref{gen}, we derive the most general solution for two domain walls in
AdS$_5$. In particular, it includes novel scenarios where the brane
tensions are not restricted to have opposite signs. The stability of
the solutions as well as generalizations to any FRW cosmologies on the
walls are discussed in section~\ref{discu}.

\section{A class of solutions in five dimensions}  \label{soln}

We are interested in two four-dimensional hypersurfaces (branes),
${\cal M}_4^{(1)}$ and ${\cal M}_4^{(2)}$, embedded in a
five-dimensional manifold such that $Z_2$ symmetry holds across each
brane. The stress-energy in the bulk is given by a negative
cosmological constant, $\lambda<0$, although, at least for part of
this section, the given solution also holds for $\lambda>0$. We shall also
assume that the energy density on each wall is dominated by a
cosmological constant (tension) denoted by $\rho_i$, $i=1,2$. With the
metric signature $(-,+,+,+,+)$ and in units where the five-dimensional
Newton constant equals unity, the action is given by 
\begin{equation}
   S=\int_{{\cal M}_5}\sqrt{-g_5}[{\cal R}_5
       -12\lambda]
     - \sum_{i=1}^2\int_{{\cal M}_4^{(i)}}d^4x\sqrt{-g_4^{(i)}}12\rho_i, 
\label{eq:action5a}
\end{equation}
where $g_4^{(i)}$ is the induced metric on the domain
walls. Without loss of generality, we can define a coordinate $y$ so
that the first domain wall is located at $y=0$. To obtain
cosmologically relevant solutions, we must impose (spatial)
homogeneity and isotropy (i.e., planar symmetry) on this
wall. Assuming that planar symmetry is also a symmetry of the bulk
(i.e., we can foliate the space-time with planar-symmetric
hypersurfaces), a general ansatz for the bulk metric is  
\begin{equation}
   ds^2_5=e^{2\beta}\left(-d\tau^2 + dy^2\right) 
      + e^{2\alpha}d\vec{x}^2 
\label{eq:ansatz}
\end{equation}
where $\alpha=\alpha(\tau,y)$ and $\beta=\beta(\tau,y)$. Note 
that we have used the freedom of coordinate reparameterization in the
$(\tau,y)$ plane to choose conformal gauge for that part of the
metric. With the assumption of planar symmetry, we shall see that the
bulk geometry must be AdS or AdS-Schwarzschild. For such bulk
geometries, the set-up is so far general, that is, one can find
coordinates in which the first domain wall is located at $y=0$. If we
want to embed a second domain wall, then, in general, its location
will be described by a function of the coordinates,
$y_2=y_2(t,\vec{x})$, such that homogeneity and isotropy on that brane
might not be manifest. However, for simplicity, we shall assume that
the second wall lies at $y=R$ (and is thus aligned with the
first). Note that by a simple change of coordinates in the $(\tau,y)$
plane, one can show that our set-up also includes cases where the
branes move uniformly according to $y_i=y_i(\tau)$. Later, we will
discuss more general configurations. 

Let us first focus our attention on the bulk solutions. One obtains
the following (bulk) equations of motion (a dot represents time
differentiation while a prime denotes differentiation with respect to
$y$) 
\begin{eqnarray}
\nonumber
& & (0,0): \;\;\;\;\;
3(\dot{\alpha}^2+\dot{\alpha}\dot{\beta}-\alpha''+\alpha'\beta'-2\alpha'^2)=6\lambda
e^{2\beta} \\
\nonumber
& & (5,5): \;\;\;\;\;
3(-\ddot{\alpha}+\dot{\alpha}\dot{\beta}-2\dot{\alpha}^2+\alpha'^2+\alpha'\beta')=-6\lambda
e^{2\beta} \\
\nonumber 
& & (i,i):
\;\;\;\;\;(2\alpha''+\beta''+3\alpha'^2)-(2\ddot{\alpha}+\ddot{\beta}+3\dot{\alpha}^2)=-6\lambda
e^{2\beta} \\
& & (0,5): \;\;\;\;\; \dot{\alpha}'+\dot{\alpha}\alpha'-\dot{\alpha}\beta'-\dot{\beta}\alpha'=0.
\end{eqnarray}
By taking linear combinations of the equations, one can
rewrite them in terms of light-cone coordinates, $x^{\pm}=\tau\pm y$,
as 
\begin{equation} 
\partial_+\partial_-\alpha+\partial_+\partial_-\beta=\frac{\lambda}{2}e^{2\beta} 
\label{eq:+-1}
\end{equation}
\begin{equation}
\partial_+\partial_-\alpha+3\partial_+\alpha\partial_-\alpha=\lambda
e^{2\beta} 
\label{eq:+-2}
\end{equation}
\begin{equation}
\partial_+^2\alpha-2\partial_+\alpha\partial_+\beta+(\partial_+\alpha)^2=0 
\label{eq:+-3}
\end{equation}
\begin{equation}
\partial_-^2\alpha-2\partial_-\alpha\partial_-\beta+(\partial_-\alpha)^2=0.
\label{eq:+-4}
\end{equation}

As shown in Ref. \cite{bine}, one can derive a first integral
of motion by rewriting the $(0,0)$ and $(5,5)$ equations as 
\begin{eqnarray}
\nonumber
& & F'(\tau,y)=-4\lambda\alpha'e^{4\alpha} \\
& & \dot{F}(\tau,y)=-4\lambda\dot{\alpha}e^{4\alpha},
\label{eq:1stint}
\end{eqnarray}
where $F(\tau,y)\equiv
e^{4\alpha}e^{-2\beta}(\alpha'^2-\dot{\alpha}^2)$. Equations
(\ref{eq:1stint}) can then be integrated to yield a single expression  
\begin{equation}
\dot{\alpha}^2-\alpha'^2=\left[\lambda+Ce^{-4\alpha}\right]e^{2\beta},
\label{eq:new}
\end{equation}
where $C$ is an integration constant. 

The assumption of planar symmetry has reduced the allowed
bulk geometries to a one-parameter family of solutions depending on
$C$. From now on we will focus on the the case $C=0$. As we will see
in the following sections, the geometry of the corresponding bulk
space-times is then particularly simple, namely AdS. However, it is
also possible to consider the case $C\neq 0$, which corresponds to a
bulk Schwarzschild-AdS geometry~\cite{ida}. 

Taking $C=0$, in light-cone coordinates, equation (\ref{eq:new}) reads
\begin{equation}
\partial_+\alpha\partial_-\alpha=\frac{\lambda}{4}e^{2\beta}.
\label{eq:+-5}
\end{equation}
We are therefore left with the $(0,5)$ equation and equation
(\ref{eq:+-5}) (since the $(i,i)$ equation then follows from the
Bianchi identity). Combining equations (\ref{eq:+-1}) and
(\ref{eq:+-2}) with equation (\ref{eq:+-5}) yields 
\begin{equation}
\partial_+\partial_-\alpha
=\partial_+\partial_-\beta=\frac{\lambda}{4}e^{2\beta}.
\label{eq:main}
\end{equation}
The solution to equations (\ref{eq:main}) and $(0,5)$ can be
expressed in terms of two arbitrary functions $f=f(x^+)$ and
$g=g(x^-)$ as 
\begin{equation}
ds^2=\frac{1}{(f+g)^2}\left(\frac{4}{\lambda}f'g'(-d\tau^2+dy^2)+d\vec{x}^2
\right),
\label{eq:bulksoln}
\end{equation}
where $f'\equiv df/dx^+$ and $g'\equiv dg/dx^-$. Although we
will be interested in the case $\lambda<0$, this solution is, in fact,
valid for any non-zero value of $\lambda$.  

Finally, to fix the specific coordinate patch on which we analyze our
solutions, let us assume that the range of $\tau$ and $y$ are such
that $\tau$ is always time-like, $y$ space-like, and further, that $f+g$
is positive. Given that $\lambda<0$ this means that we will take
\begin{equation}
 f(x^+) + g(x^-) > 0 
\label{eq:fgcond1}
\end{equation}
\begin{equation}
f'(x^+)g'(x^-) < 0 .
\label{eq:fgcond2}
\end{equation}
for all $x^+$ and $x^-$. 

Let us now turn to the domain walls. We must 
supplement the equations of motion with appropriate boundary
conditions. Since the branes are localized objects along the
transverse direction (delta-function sources), they result in
discontinuities in the normal derivative of the metric. Formally,
these discontinuities are described by the Israel matching
conditions~\cite{israel, chamblin}, which yield the following boundary
conditions (assuming $Z_2$ symmetry across each 
wall) 
\begin{equation}
   e^{-\beta}\alpha'|_{y=0} = - \rho_1, \qquad
   e^{-\beta}\alpha'|_{y=R} = + \rho_2, 
\label{eq:bc1}
\end{equation}
\begin{equation}
   e^{-\beta}\beta'|_{y=0} = -\rho_1, \qquad
   e^{-\beta}\beta'|_{y=R} = +\rho_2.
\label{eq:bc2}
\end{equation}
where we note that there is a difference in sign between the two walls. 

The Israel junction conditions impose restrictions on the functions
$f$ and $g$. From the boundary condition (\ref{eq:bc1}) evaluated at
$y=0$ and given the conditions~(\ref{eq:fgcond1}) and (\ref{eq:fgcond2}), we obtain 
\begin{eqnarray}
\nonumber
   && \qquad g'(\tau)-f'(\tau) =
      -\rho_1\sqrt{\frac{4}{\lambda}f'(\tau)g'(\tau)}  \\ 
   && \Longrightarrow \left(-\frac{g'(\tau)}{f'(\tau)}\right)^{1/2} = 
      \frac{1}{\sqrt{-\lambda}}(\rho_1\pm H_1)\equiv \gamma_1,
\label{eq:bd1}
\end{eqnarray}
where we have introduced
\begin{equation}
   H_i = \sqrt{\rho_i^2 + \lambda}.
\label{eq:H}
\end{equation}
Since $H_i$ should be real we find that solutions will only exist if we
have the condition on the tension,
\begin{equation}
   \label{eq:tension}
   |\rho_i| \geq \sqrt{-\lambda}.
\end{equation}
Note we are using here the condition~(\ref{eq:fgcond2}), which implies
that $g'(x^-)/f'(x^+)<0$ for all $x^+$ and $x^-$. It turns out that the
relation~(\ref{eq:bd1}) then implies that the second boundary condition
(\ref{eq:bc2}) is also satisfied at $y=0$.  

At $y=R$, the boundary conditions imply a similar relationship between
$g(\tau)$ and $f(\tau+2R)$. One finds that 
\begin{equation}
   \label{eq:bd2}
   \left(-\frac{g'(\tau)}{f'(\tau+2R)}\right)^{1/2} = 
      \frac{1}{\sqrt{-\lambda}}(-\rho_2\pm H_2)\equiv \gamma_2
\end{equation}
satisfies both boundary conditions. 

Together, equations~(\ref{eq:bd1}) and~(\ref{eq:bd2}) imply that 
\begin{eqnarray}
   && g(\tau) = - \gamma_1^2 f(\tau)-k_1, \nonumber \\
   && g(\tau) = - \gamma_2^2 f(\tau+2R)-k_2,  
\label{eq:fg}
\end{eqnarray}
where $k_i$ are constants and 
\begin{eqnarray}
   && \gamma_1 \equiv \frac{1}{\sqrt{-\lambda}}(\rho_1 \pm H_1) ,
      \nonumber \\
   && \gamma_2 \equiv \frac{1}{\sqrt{-\lambda}}(-\rho_2 \pm H_2) .
\label{eq:gammai}
\end{eqnarray}
Note that, by~(\ref{eq:bd1}) and~(\ref{eq:bd2}) both $\gamma_i$ must be
positive. Since $|\rho_i|>H_i$ this requires
\begin{equation}
   \label{eq:rho}
   \rho_1 > 0, \qquad \rho_2 < 0,
\end{equation}
that is, for solutions of this type, the brane tensions must have
opposite sign. As we shall see in section~\ref{gen}, this condition on
the brane tensions is a consequence of our particular choice of
coordinate system (\ref{eq:ansatz}). We will later also describe
solutions where this condition does not hold.   

Together, the relationships~(\ref{eq:fg}) imply a
periodicity condition on $f$ which can be written as
\begin{equation}
   f(x+2R) = \left(\gamma_1/\gamma_2\right)^2 f(x) + K,
\label{eq:period}
\end{equation}
where the constant $K$ is given by
\begin{equation}
   K = (k_1-k_2)/\gamma_2^2 
\end{equation}
The general solution of this periodicity condition gives
\begin{equation}
   \label{eq:periodsol}
   f(x) = e^{-\xi x}p(x) + \frac{e^{-\xi x}-1}{e^{-2\xi R}-1}K,
\end{equation}
where
\begin{equation}
   \label{eq:xi}
   \xi \equiv -\frac{1}{R}\ln(\gamma_1/\gamma_2),
\end{equation}
and $p(x)$ is a periodic function $p(x+2R)=p(x)$. One notes that in
the limit $\xi\rightarrow 0$ the general expression~(\ref{eq:periodsol}) becomes
\begin{equation}
   f(x) = p(x) + \frac{Kx}{2R}.
\end{equation}

In the following, we will sometimes find it easier to use
$k\equiv-k_1$ and $K$ as the independent constants in our solution,
and will sometimes stick to $k_1$ and $k_2$. For completeness let us
now give the general expressions for the functions $\alpha$ and
$\beta$ in the metric in terms of $f$. We have
\begin{eqnarray}
   && e^\alpha = 
       \frac{1}{f(x^+) - \gamma_1^2 f(x^-) + k}, \nonumber \\
   && e^\beta 
       = \sqrt{\frac{4\gamma_1^2f'(x^+)f'(x^-)}{-\lambda}}\; e^\alpha. 
\label{eq:ab}
\end{eqnarray}

We see that the general solution is determined by the
choice of a periodic function $p(x)$ and some constants, which must be
chosen so that equations~(\ref{eq:fgcond1}) and (\ref{eq:fgcond2}) are satisfied. While it is
difficult at this point to get a feel for what 
these parameters represent, we will find in sections~\ref{dynamics}
and~\ref{l<0} a geometrical interpretation for
them. Note that our solution reduces to the solution 
presented in Ref. \cite{kim} if we choose $p={\rm constant}$ (as we
shall see later, such a choice amounts to a coordinate
transformation). Nevertheless, we shall find evidence in section
\ref{l<0} (and show in section \ref{gen}) that the solution in this
section is not the most general.  

Let us now consider the induced geometry on the branes. A simple
approach is to substitute the boundary conditions (\ref{eq:bc1}) and
(\ref{eq:bc2}) in equation (\ref{eq:new}) (with $C=0$) to find the
Friedmann equation on the domain wall following~\cite{bine}. Choosing
cosmological time on the brane, one finds 
\begin{equation}
\dot{\alpha}^2=\rho_i^2+\lambda = H_i^2.
\label{eq:binetruy}
\end{equation}
Given the condition on the tensions~(\ref{eq:tension}), equation
(\ref{eq:binetruy}) implies that the induced geometry on each 
brane is precisely that of de Sitter space with cosmological constant
$H_i$. 

We can also see this more explicitly from our final
expression~(\ref{eq:ab}). Considering the brane at
$y=0$ for instance, we can perform the following time redefinition  
\begin{equation}
   e^{-H_1t} = (1-\gamma_1^2) f(\tau) + k 
\label{eq:transfn}
\end{equation}
so that $t$ is cosmological time at $y=0$. In terms of the
new time variable $t$, the induced metric on our domain wall is found
to be  
\begin{equation}
   ds_4^2=-dt^2+e^{2H_1t}d\vec{x}^2,
\label{eq:4dmetric}
\end{equation}
which describes de Sitter space, in agreement with equation
(\ref{eq:binetruy}). A similar analysis holds for the brane at $y=R$. 

To conclude this section, let us discuss an important subclass of
solutions, namely those with a static orbifold in the coordinates
of~(\ref{eq:ansatz}). Starting with the case 
$\rho_1=-\rho_2$ (i.e., $\xi=0$), we see from equations 
(\ref{eq:ab}) that a static bulk is obtained only if
$\gamma_1^2=1$ and $p={\rm constant}$. We will show in
section~\ref{l<0} that this solution corresponds to the
Randall--Sundrum scenario \cite{randall}. 

In the case $\rho_1\neq -\rho_2$ (i.e., $\xi\neq 0$), equations
(\ref{eq:ab})  tell us that static solutions require $p={\rm
  constant}$ and $k=\frac{(1-\gamma_1^2)K}{e^{-2\xi R}-1}$. For such choices, one
finds the following constraint relating the physical distance between
the walls, $R_{static}$, the bulk cosmological constant and the brane tensions 
\begin{equation}
\tanh(\sqrt{-\lambda}R_{static}) =
   \frac{\sqrt{-\lambda}(\rho_1+\rho_2)}{-\lambda+\rho_1\rho_2},  
\label{eq:statickal}
\end{equation}
in agreement with Refs.\cite{kaloper} and
\cite{nihei}. Since $\rho_i^2+\lambda\geq 0$ and $\rho_1\rho_2<0$, the
denominator in equation (\ref{eq:statickal}) is negative. In turn,
this implies that $-\rho_2>\rho_1>0$.

\section{Geometrical analysis of $\lambda=0$ solution} \label{dynamics}

In this section, we shall discuss the limit $\lambda\rightarrow 0^-$
of our solution from a geometrical perspective. In the process, we
will find geometrical interpretations for the various parameters
describing the general solution (such as $k$, $K$, etc.). 

Let us first determine the bulk geometry in this limit. We note that
general solutions with a metric in the form~(\ref{eq:ansatz}) with
$\lambda=0$ were first discussed (albeit in four dimensions) by
Taub~\cite{taub} (see also~\cite{ipser}). In that paper it was shown
that all solutions with $C=0$ were simply different parametrizations
of flat space, while all solutions with $C\neq 0$ were simply
different parametrizations of the Kasner ``rolling radii''
solutions~\cite{kasner}. Here we will rederive the $C=0$ case as a
limit of our general solution. From equation (\ref{eq:bulksoln}), we
note that taking $\lambda\rightarrow 0$ requires letting either $f'$
or $g'$ go to zero at the same rate. Suppose we let $g'\rightarrow
0$. Defining  
\begin{equation}
   g(x^-) = \frac{\lambda}{4M^2}\tilde{g}(x^-)+\tilde{k},
\label{eq:tildeg}
\end{equation}
where $\tilde{k}$ is a constant and $M$ some energy scale, equation
(\ref{eq:bulksoln}) becomes 
\begin{equation}
   ds^2 = \frac{1}{(f(x^+)+\tilde{k})^2}
        \left(-\frac{1}{M^2}df(x^+)d\tilde{g}(x^-)+d\vec{x}^2\right).
\end{equation}
To express this metric in a more familiar form, we can perform the
coordinate transformation to $(T,Y,\vec{X})$
\begin{eqnarray}
\nonumber
  && Y+T = \frac{1}{M(f(x^+)+\tilde{k})} \\
\nonumber
  && Y-T =  \frac{1}{M}\tilde{g}(x^-) - \frac{M\vec{x}^2}{f(x^+)+\tilde{k}} \\
  &&\quad\;\;\;
   \vec{X} = \frac{\vec{x}}{f(x^+)+\tilde{k}}, 
\label{eq:coordchange}
\end{eqnarray}
which gives
\begin{equation}
   ds^2 = -dT^2+dY^2+d\vec{X}^2.
\label{eq:flat}
\end{equation}
Thus we have shown that the case $C=0$
case indeed corresponds to five-dimensional Minkowski space 
(${\cal M}_5$) in the limit $\lambda\rightarrow 0$. If we now consider the 
branes at the orbifold fixed points, this imposes boundary conditions
on $f$ and $g$ (see eqns. (\ref{eq:fg})). For these to be consistent
with equation (\ref{eq:tildeg}), we need to choose the signs in
$\gamma_i$ (see eq. (\ref{eq:gammai})) such that $\gamma_i\rightarrow
0$ as $\lambda\rightarrow 0$ (note also that this yields
$\gamma_1/\gamma_2\rightarrow |\rho_2/\rho_1|$). Furthermore, we must 
identify $\tilde{k}$ with the constant $k$ introduced in section
\ref{soln}, $M$ with the tension $\rho_1$ of the $y=0$ brane, and
$\tilde{g}(x^-)=f(x^-)$ (from eq. (\ref{eq:fg})). Similarly, the above
derivation can be repeated for the choice $f'\rightarrow 0$. Note that
our solution in the limit $\lambda\rightarrow 0$ agrees with the
results presented in Ref. \cite{lukas}. Finally, combining equation~(\ref{eq:fgcond1}) with the first of equations~(\ref{eq:coordchange}), we note that our original coordinates are mapped to the range $Y+T>0$.

\begin{figure}[ht]
\begin{center}
\includegraphics{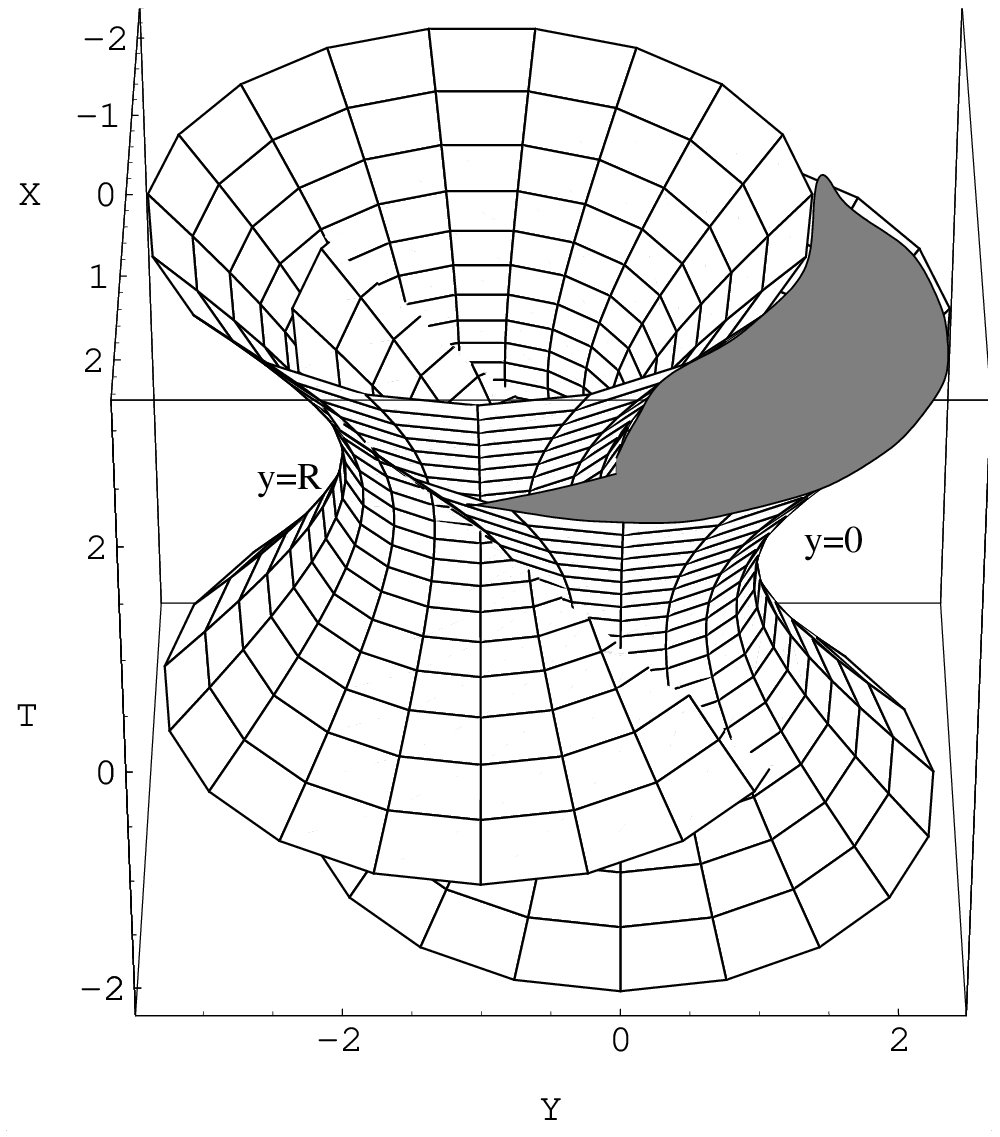}
\end{center}
\caption{Illustration of the $\rho_1=-\rho_2$ solution with parameters
  $k=0$, and $K>0$ (note that two spatial dimensions have
  been suppressed) in the case where the bulk is flat space. The
  location of the bulk (indicated by the shaded region) is determined
  by the curvatures of the boundaries. While the hyperboloids are seen
  to intersect, they do so on the plane $Y+T=0$, a surface which can
  never be reached by an observer in the shaded region.}  
\label{fig:K3D}
\end{figure}

We saw in the previous section that the induced geometry on each domain
wall is that of de Sitter space. We
can therefore interpret the solutions obtained in section
\ref{soln} (for $\lambda\rightarrow 0$) as possible embeddings of two dS$_4$ surfaces in ${\cal
M}_5$. As in well-known (e.g., see Ref. \cite{hawking}), a de Sitter surface
can be described as a hyperboloid embedded in ${\cal M}_5$. Indeed,
evaluating equations (\ref{eq:coordchange}) at $y=0$ and $y=R$, one
finds that the branes are given by  
\begin{eqnarray}
\nonumber
& &
   - \left(T-\frac{k}{2\rho_1}\right)^2 
   + \left(Y+\frac{k}{2\rho_1}\right)^2
   + \vec{X}^2 = \frac{1}{\rho_1^2} \qquad \qquad \qquad \qquad
   {\rm for} \quad y=0, \\
& & 
   - \left(T-\frac{\rho_1(k+K)}{2\rho_2^2}\right)^2
   + \left(Y+\frac{\rho_1(k+K)}{2\rho_2^2}\right)^2
   + \vec{X}^2 = \frac{1}{\rho_2^2} \qquad
   {\rm for} \quad y=R.
\label{eq:hyper1}
\end{eqnarray}
From equations (\ref{eq:hyper1}), we note that:
\begin{itemize}
\item The curvature of the hyperboloids are given by
$|\rho_i|$.
\item The origins of the light cones asymptotic to the hyperboloids
   are separated by the {\it null vector} 
   \begin{equation}
      \label{eq:nullvector}
      D = \left[(\rho_1/2\rho_2^2)(K+k)-k/2\rho_1\right]\cdot(-1,1,\vec{0}).
   \end{equation}
\item For fixed null separation $D$, the quantity $k$ simply translates the two hyperboloids by the
   same vector. It therefore describes the invariance of the solution
   upon translations in the embedding space and corresponds to a
   change of coordinates in~(\ref{eq:flat}).
\end{itemize}
While the intrinsic curvature of each hyperboloid is given by
$|\rho_i|$, the sign of $\rho_i$ determines on which side of the
hyperboloid the bulk extends. The bulk lies outside (inside) the
hyperboloid if $\rho_i<0$ ($>0$). As an example, figure \ref{fig:K3D}
shows the solution with $\rho_1=-\rho_2$ and parameters $k=0$ and
$K>0$, (where we have suppressed two spatial dimensions). In
accordance with the above discussion, the portion of space-time 
consistent with $\rho_1>0$ and $\rho_2<0$ is indicated by the shaded
region. From~(\ref{eq:hyper1}) we see that the two hyperboloids
intersect in the cylinder $Y+T=0$, $\vec{X}^2=1/\rho_1^2$. Since
$Y+T=0$ is a null plane and given our initial coordinate patch, it is
easy to show that no observer in the shaded region can ever reach this
intersection. Due to the symmetry in the $\vec{X}$ directions, it is
sufficient for our purposes to focus on the two-dimensional projection
onto the $\vec{X}=0$ plane (shown in fig. \ref{fig:hyper1} for the
above example). An important point to notice is that the hyperbolas
are separated by a null vector (as mentioned in the second point
above), and therefore share a common asymptote.

We can also consider examples where the curvatures of the hyperboloids
are not equal, once again focusing on the subspace $\vec{X}=0$.
 For simplicity, let us set $k=0$. The case $K=0$ with $-\rho_2>\rho_1>0$
is shown in figure \ref{fig:hyper1b}, with the bulk indicated by the
shaded region. Note that the branes never intersect for $K=0$. Furthermore, 
it is crucial that $-\rho_2>\rho_1$ in order to obtain a bulk region that is simultaneously
inside one brane and outside the other, which is equivalent 
to the condition stated below equation (\ref{eq:statickal}). 

\begin{figure}[ht]
\begin{center}
\includegraphics{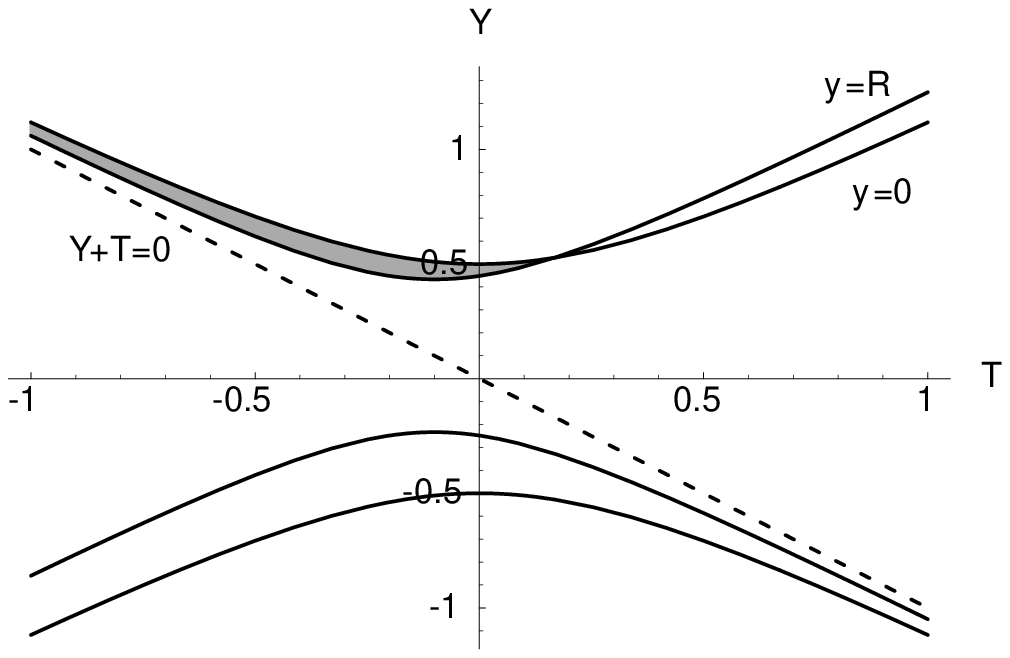}
\end{center}
\caption{Illustration of the solution with $-\rho_2>\rho_1>0$ and $K<0$ (and thus, $B>0$).}
\label{fig:hyper2}
\end{figure}

\begin{figure}[ht]
\begin{center}
\includegraphics{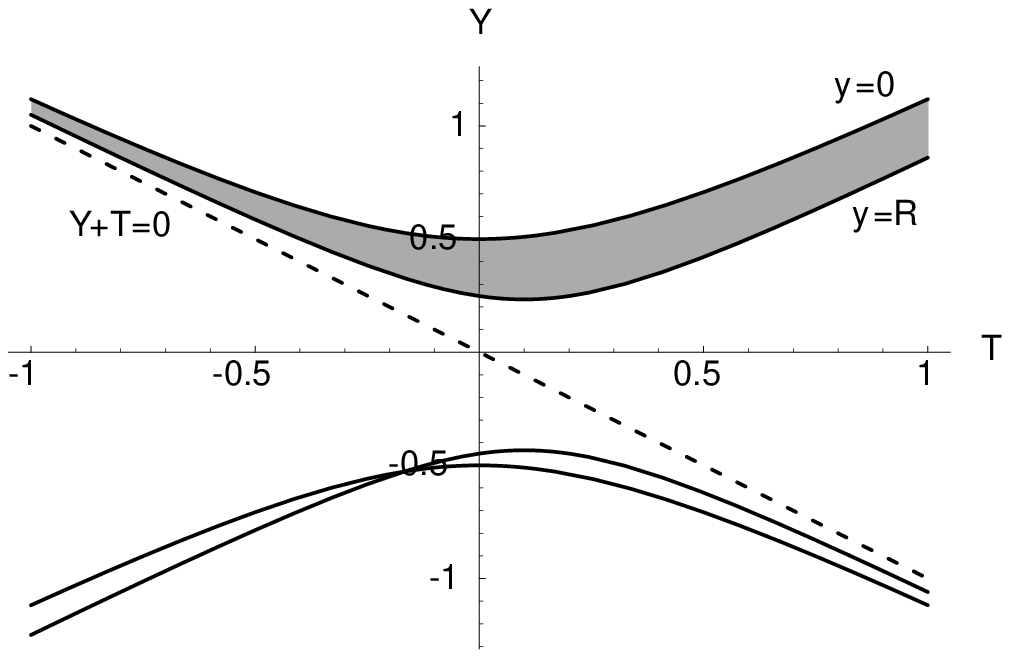}
\end{center}
\caption{Illustration of the solution with $-\rho_2>\rho_1>0$ and $K>0$ ($B<0$). }
\label{fig:hyper2a}
\end{figure}

For the case $K\neq 0$, we see from~(\ref{eq:hyper1}) that the
hyperboloids now do intersect in the paraboloid
\begin{eqnarray}
   \label{eq:intersect}
   \nonumber
   && Y+T = B 
      \equiv \frac{\rho_2^{-2}-\rho_1^{-2}}{(\rho_1/\rho_2^2)(K+k)-k/\rho_1} \\
   && Y-T = -\frac{1}{B}\vec{X}^2 + \frac{1}{B\rho_1^2} - \frac{k}{\rho_1}.
\end{eqnarray}
Note that, as above, the intersection lies in a null plane. Furthermore,
since our coordinates are restricted to $Y+T>0$, the intersection lies within our
coordinate patch only if $B>0$. The solution for $-\rho_2>\rho_1>0$
with $K<0$ and $K>0$ are illustrated in figures \ref{fig:hyper2} and
\ref{fig:hyper2a}, respectively. These two figures are related by reflections 
about the $T$ and $Y$ axes, but they should be viewed as distinct solutions in our
analysis due to the constraint $Y+T>0$. In particular, the intersection 
lies in the physically accessible region in figure~\ref{fig:hyper2}, where $B>0$, 
but not in figure~\ref{fig:hyper2a}, where $B<0$. Finally, if $\rho_1>-\rho_2$, 
the only choice consistent with there being a bulk between the two branes in the 
region $Y+T>0$ is $K>0$, in which case there is also an intersection in our 
coordinate patch, as illustrated in figure~\ref{fig:hyper2b}. (For the case $K<0$, 
the signs of the brane tensions do not allow for a bulk region between them in 
the region $Y+T>0$.)

We have shown that a broad class of configurations can be embedded in
flat space, namely those consisting of null-separated de Sitter
surfaces. The fact that we have only encountered cases with null
separation is surprising, and leads us to question the generality of
the solutions obtained in section \ref{soln}. Intuitively, one would
also expect cases with time-like and space-like separated
hyperboloids. We shall confirm this intuition in section \ref{gen}. 

\subsection{Dynamics of the orbifold}

We next want to investigate the evolution of the orbifold as seen by
an observer on the brane at $y=0$. First, let the cosmological time
for this observer be denoted by $t^{(1)}$ (i.e., $t^{(1)}$ puts the induced
metric at $y=0$ in the form given in eq. (\ref{eq:4dmetric})). The
range $-\infty<t^{(1)}<\infty$ is mapped to the range
$-\infty<T<\infty$, with the rule that $t^{(1)}$ increases with $T$ in
the region $Y+T>0$. Note that requiring the time coordinate to be
cosmological time at $y=0$ does not uniquely fix the surfaces of
constant time away from the brane. In this sense, the distance between
the branes is not a coordinate-independent notion. However, a natural
prescription is the following. First, choose coordinate such that the
metric has the form~(\ref{eq:ansatz}). Then recall that we have seen
that that neither the function $p(x)$ nor $k$ in the solution encode
physical information. Thus it is natural to take the case 
$p={\rm const}$ and $k=0$. One then denotes the separation as the distance in the
$y$-direction with fixed $t$ and $\vec{x}$ in this metric.

\begin{figure}[ht]
\begin{center}
\includegraphics{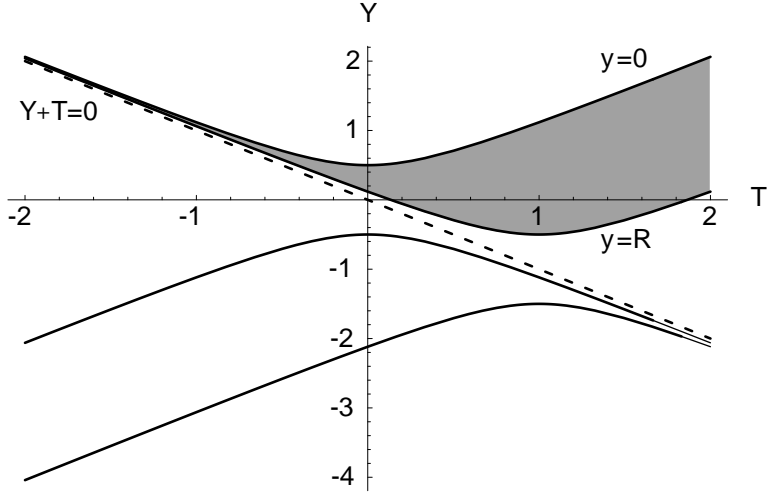}
\end{center}
\caption{Two-dimensional ($\vec{X}=0$) projection of the previous
  figure (with $k=0$, and $K>0$). The bulk (as determined by
  the curvature of the domain walls) is indicated by the shaded
  region. Note that the bulk region satisfies the requirement that it is 
above the dotted line $Y+T=0$, as required for it to be within the physically accessible coordinate 
patch (defined by $Y+T>0$).} 
\label{fig:hyper1}
\end{figure}

\begin{figure}[ht]
\begin{center}
\includegraphics{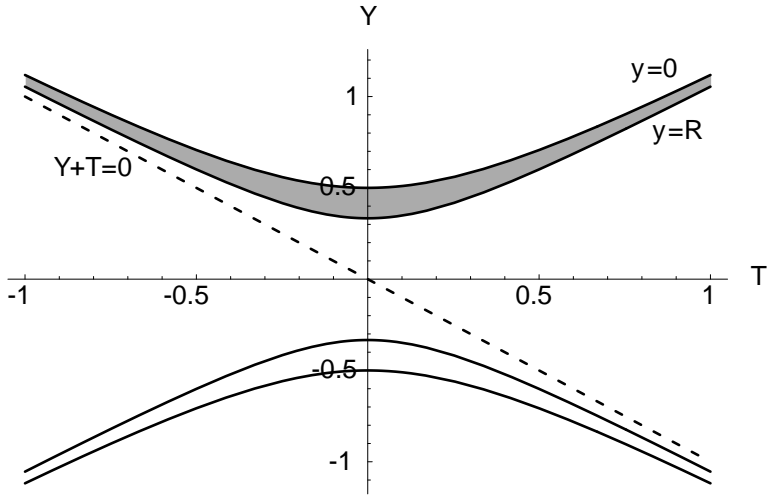}
\end{center}
\caption{Illustration of the solution with $-\rho_2>\rho_1>0$ and $K=0$ projected 
onto the $\vec{X}=0$ plane (corresponding to $B=0$). Once again, the bulk 
corresponds to the shaded region. While the lower half of the figure could 
also yield a consistent bulk region, it lies outside the range of our original coordinates.}
\label{fig:hyper1b}
\end{figure}

\begin{figure}[ht]
\begin{center}
\includegraphics{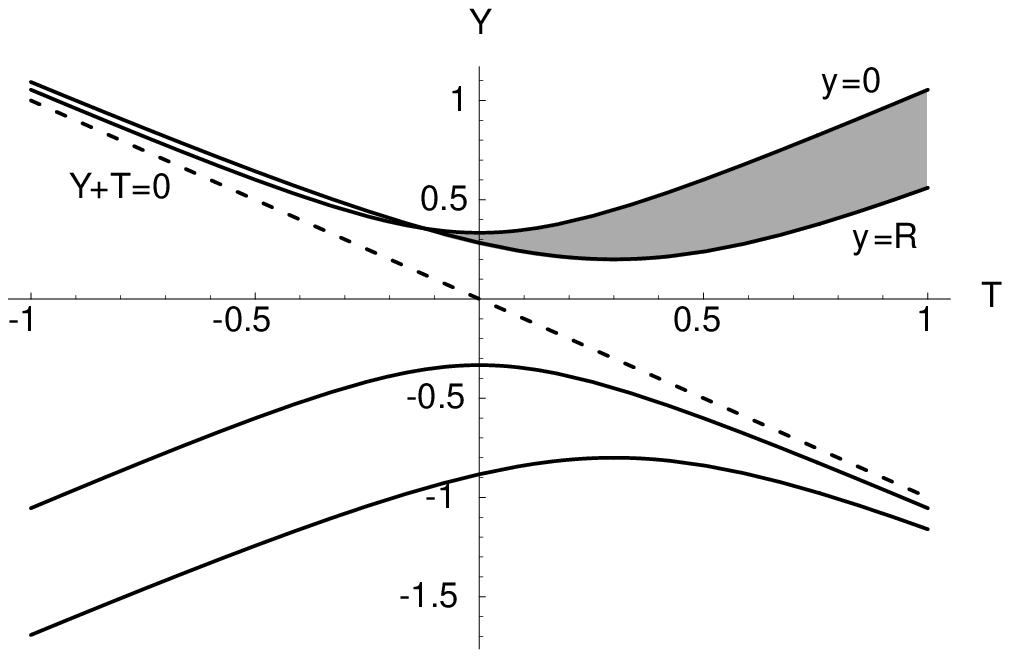}
\end{center}
\caption{Illustration of the solution with $0<-\rho_2<\rho_1$ and $K>0$ ($B>0$).}
\label{fig:hyper2b}
\end{figure}

As seen in figures \ref{fig:hyper1}-\ref{fig:hyper2b}, we have five
physically distinct solutions to consider, which we classify into
three cases.  
\begin{itemize}
\item {\bf $-\rho_2>\rho_1$}:
\begin{itemize}
\item $K=0$: This solution is shown in figure \ref{fig:hyper1b}. As
   mentioned at the end of section \ref{soln}, it has a static orbifold
   due to the choice $p=const$. The physical length of the orbifold is
   given by the $\lambda\rightarrow 0$ limit of equation
   (\ref{eq:statickal}), namely
\begin{equation}
R_{static}\rightarrow\frac{1}{\rho_1}+\frac{1}{\rho_2},
\end{equation}
and agrees with the expression found in Ref. \cite{lukas}. Finally,
the curves of constant $t^{(1)}$ are straight lines trough the
origin. 
\item $K<0$: This solution corresponds to the shaded region in
   fig. \ref{fig:hyper2}. If we denote the physical distance between
   the branes by $R_{phys}$ in this case, one finds that
   $R_{phys}\rightarrow R_{static}^-$ as $t^{(1)}\rightarrow
   -\infty$. As $t^{(1)}$ increases, the distance between the domain
   walls decreases until it reaches zero size (corresponding to the
   point where the hyperboloids intersect). Hence, if we define the
   onset of inflation to occur at at some $t^{(1)}_0$, where
   $-\infty<t^{(1)}_0<t^{(1)}_{collapse}$, this solution describes an
   initial condition where the branes are within $R_{static}$ apart,
   and subsequently move toward each other. 
\item $K>0$: In this case (shown in fig.~\ref{fig:hyper2a}),
   $R_{phys}\rightarrow R_{static}^+$ as $t^{(1)}\rightarrow
   -\infty$. Subsequently, $R_{phys}$ increases forever.  
\end{itemize}
Hence, for the case $-\rho_2>\rho_1$, the orbifold remains static if
the initial distance is $R_{static}$, collapses if it is less than
$R_{static}$, and greater if it is larger than $R_{static}$. This
conclusion agrees with the analysis of Ref. \cite{kim}.  
\item {\bf $-\rho_2<\rho_1$}: This solution was not discussed in
   Ref. \cite{kim}. From figure \ref{fig:hyper2b}, we see that it
   describes an orbifold starting from zero size and expanding to
   infinity. Hence, in this case, any initial distance leads to
   expansion. 
\item {\bf $-\rho_2=\rho_1$}: By comparing figures \ref{fig:hyper1}
   and \ref{fig:hyper2b}, we see that the evolution is the same as in
   the case $-\rho_2<\rho_1$.  
\end{itemize}

As mentioned earlier, different choices of $p$ correspond to different
observers in the sense that they are associated with diffeomorphisms
which leave the bulk metric in the form~(\ref{eq:ansatz}) and the
induced metric at $y=0$ in the form~(\ref{eq:4dmetric}). The notion of
the separation of the branes is coordinate dependent. To see this
explicitly suppose, for instance, we allow for a generic choice of
$p$. It turns out that an observer at $y=0$ will then see, on top of
the average behavior of the extra dimension, an oscillatory
component. For instance, in the case $-\rho_2<\rho_1$, the orbifold
would oscillate as it expands. Note that, given the restrictions
imposed by equations (\ref{eq:fgcond1}) and (\ref{eq:fgcond2}), the function $p$ must be
chosen such that its oscillations will not lead to $R_{phys}$ reaching
zero size where it would not in the case $p=const$. 

Nonetheless, there is also an observer-independent statement one can
make about the various solutions described above. We noted that, in
general, the branes intersect in a paraboloid~(\ref{eq:intersect}). 
This intersection lies in the coordinate patch $Y+T>0$ if $B>0$. 
Furthermore, it lies a finite distance in the future of any observer 
in the bulk (corresponding to an orbifold collapsing to a final singularity) 
if $-\rho_2>\rho_1>0$, or in the past (corresponding to an orbifold expanding 
from an initial singularity) if $\rho_1>-\rho_2>0$. Finally, let us note that 
the surface of intersection is formally a singular surface and, thus, our 
solution breaks down in its neighborhood. A full treatment of brane collisions 
would require including effects due to the finite thickness of the branes 
(usually of string size).

\subsection{Causal properties}

Let us now consider the causal properties of the solution. More
precisely, we want to see if the walls are causally-connected during
the inflationary period, and whether two-way or one-way communication
between the walls is possible.  

In the cases where the orbifold collapses or remains static
(i.e. $K\leq 0$), two-way communication is possible for all times, as
expected.  

\begin{figure}[ht]
\begin{center}
\includegraphics{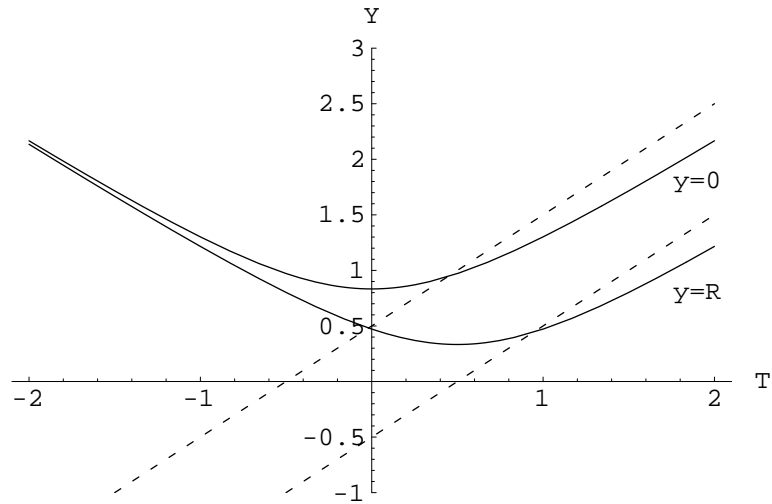}
\end{center}
\caption{Same case as that of figure \ref{fig:hyper1} except that the
  dotted lines now represent gravitons sent from $y=R$ towards
  $y=0$. This shows that after a while, signals sent from $y=R$ do not
  reach $y=0$ in finite affine parameter.} 
\label{fig:hyper3}
\end{figure}

The remaining solutions involve expanding orbifolds. Since the extra
dimension is in fact inflating, one would expect that a signal sent
from one brane could never reach the other. However, we must keep in
mind that the length of the orbifold is computed along a space-like
path, while gravitons propagate along null geodesics. Consequently,
the causal properties are not as trivial as one would have
guessed. The analysis yields qualitatively the same result for all
solutions which describe expanding orbifolds, so, without loss of
generality, we can focus on the case $\rho_1=-\rho_2$ for
concreteness. Suppose observer 2 at $y=R$ sends signals to observer 1
at $y=0$. The two dotted lines in figure \ref{fig:hyper3} correspond
to two such signals. We see that the first signal reaches $y=0$ in
finite affine parameter while the other does not. Hence, from the point of
view of observer 2, there exists some time on his clock where his
signals stop reaching the $y=0$ brane. We shall say that a {\it
causal boundary} has appeared for observer 2. Now, consider signals sent from
$y=0$ towards $y=R$. It is clear from figure \ref{fig:hyper3} that
such signals always reach $y=R$ in finite affine parameter. 
{\it However}, if $t^{(1)}$ is physical time for observer 1, then
after a while observer 1 sees his signals taking infinite time
$t^{(1)}$ to reach $y=R$. From the point of view of observer 1, we
shall say that there is a {\it horizon} somewhere in the bulk. Note
that this is not a genuine horizon in the five-dimensional space-time,
but corresponds to a surface with the following property: once a
signal sent by observer 1 has passed through the horizon, it cannot
return to observer 1. Note that causality implies that the appearance
of a horizon requires the appearance of a causal boundary, and vice
versa. To see this, suppose the contrary is true; that is, suppose
that there is a horizon for observer 1 but no causal boundary for
observer 2. If observer 1 sends a graviton towards $y=R$, because of
the horizon he will see his graviton frozen somewhere in the bulk. On
the other hand, observer 2 will receive this signal within finite time
on his clock. If observer 2 replies to the signal, his reply will be
received at $y=0$ in finite time on observer 1's clock (because we
have assumed that there were no causal boundary). From the point of
view of observer 1, he received a reply from observer 2 before his
initial signal made it to $y=R$, an obvious violation of causality. It
is easily seen that the horizon and causal boundary describe the same
surface in the bulk, namely the future light cone asymptotic to the
$y=0$ hyperboloid. That is, once observer 2 is within this surface,
his signals will not reach $y=0$ in finite affine parameter.  

To summarize our conclusions for expanding extra dimensions, our
solution predicts that two-way communication  is only possible for a
finite amount of time. Afterwards, only the brane with positive
tension can send signals  to (and, hence, have influence on) the
negative-tension brane.

\section{Geometrical analysis of $\lambda<0$ solution} \label{l<0}

In this section, we generalize the analysis of the previous section to
the case of negative bulk cosmological constant. As there, we will find
that all the solutions presented in section~\ref{soln} correspond to
the same bulk space-time. In this case it is AdS$_5$. We can therefore
think of our solution as a class of possible ways to embed two dS$_4$
surfaces in AdS$_5$. As in the case of flat space discussed in the
previous section, we shall find evidence that our solution does not
describe the full spectrum of such embeddings.   

Recall that the general bulk solution has the form~(\ref{eq:bulksoln})
\begin{equation}
   ds^2 = \frac{1}{(f+g)^2}\left(
      \frac{4}{\lambda}f'g'(-d\tau^2 + dy^2) + d\vec{x}^2 \right),
\end{equation}
With the line element in this form, it is not clear that, in fact, for
$\lambda<0$, in all cases the bulk geometry is that of
AdS$_5$. However, upon the coordinate transformation  
\begin{eqnarray}
   && z + t = \frac{2}{\sqrt{-\lambda}}f \nonumber \\
   && z - t = \frac{2}{\sqrt{-\lambda}}g,
\label{eq:coord}
\end{eqnarray}
we have
\begin{equation}
   ds^2 = \frac{1}{(-\lambda)z^2}(-dt^2+d\vec{x}^2+dz^2),
\label{eq:bulk}
\end{equation} 
which is a more familiar form for the line element of AdS$_5$. We see
that the general functions $f$ and $g$ simply represented different
choices of conformal coordinates $\tau$ and $y$ in the AdS$_5$ space. 

\begin{figure}[ht]
\begin{center}
\includegraphics{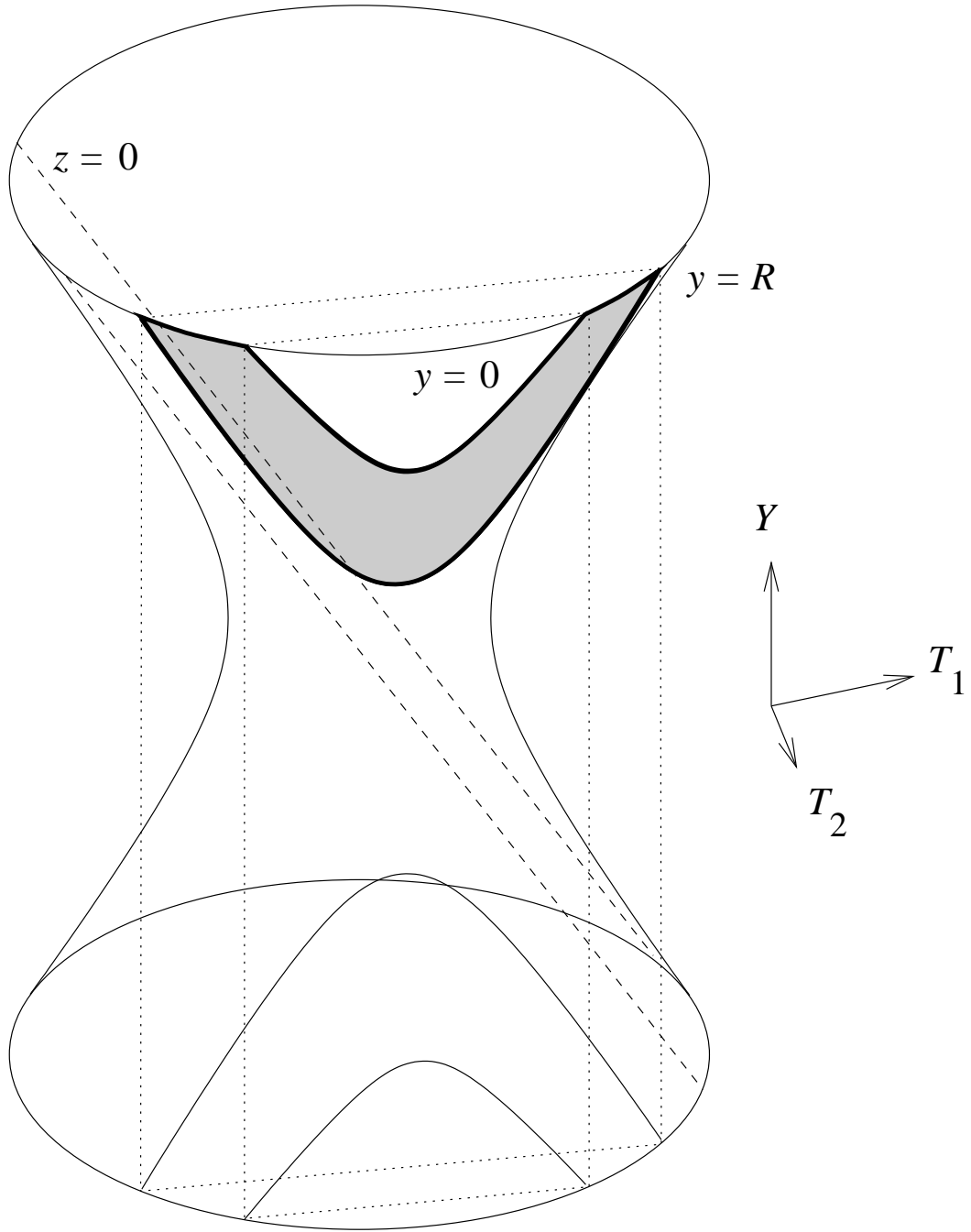}
\end{center}
\caption{Sketch of the solution with $\lambda<0$, $-\rho_2>\rho_1>0$,
  and $K=0$. The resulting space-time (indicated by the shaded region)
  is obtained by intersecting the AdS$_5$ hyperboloid with two planes
  of constant $T_2$.} 
\label{fig:ads}
\end{figure}

As is well known, this coordinate system does not cover the whole of
AdS space. More generally one can describe AdS as a
hyperboloid~\cite{hawking} 
\begin{equation}
   - X^2 \equiv T_1^2 + T_2^2 - Y^2 - \vec{X}^2 = \frac{1}{-\lambda}
\label{eq:adshyper}
\end{equation}
embedded in a flat six-dimensional space $X=(T_1,T_2,Y,\vec{X})$ with
line element  
\begin{equation}
   ds^2= dX\cdot dX \equiv  - dT_1^2 - dT_2^2 + dY^2 + d\vec{X}^2.
\end{equation}
We should note here that the space described contains closed time-like
curves. The full AdS space is really the universal covering
space. This will not enter our analysis here, since in all our
solutions the domain walls will lie in the same sheet of the universal
cover. In terms of the coordinates in~(\ref{eq:bulk}) the embedding is
given by
\begin{eqnarray}
  && T_1 + Y = \frac{1}{(-\lambda)z} \nonumber \\
  && T_1 - Y = \frac{\vec{x}^2-t^2+z^2}{z} \nonumber \\
  && T_2 = \frac{t}{\sqrt{-\lambda}z} \nonumber \\
  && \vec{X}=\frac{\vec{x}}{\sqrt{-\lambda}z}.
\end{eqnarray}
Note that since $z=(f+g)/\sqrt{-\lambda}>0$, our original coordinates
are mapped to the range $T_1+Y>0$. (The boundary of this region,
$z=0$, is shown in figure~\ref{fig:ads}.)

Let us now turn to the description of the domain walls. Recall that in
our original coordinates the walls were fixed at $y=0$ and
$y=R$, or equivalently $x^+=x^-$ and $x^+=x^-+2R$. After the conformal
transformation~(\ref{eq:coord}) the walls will now be \textit{moving}
in the $z$ direction. Explicitly, from the relations~(\ref{eq:fg}),
the equations of the walls in terms of $z$ and $t$ are the linear relations
\begin{eqnarray}
   & \displaystyle 
     t-z = \gamma_1^2 (t+z) + \frac{2k_1}{\sqrt{-\lambda}} &
        \qquad \textrm{for $y=0$}; \nonumber \\
   & \displaystyle
     t-z = \gamma_2^2 (t+z) + \frac{2k_2}{\sqrt{-\lambda}} &
        \qquad \textrm{for $y=R$}.
\label{eq:moving}
\end{eqnarray}
It is easy to see that in these coordinates, the domain walls
move in the $z$ direction with constant velocity $H_i/|\rho_i|$.

Finally we can transform these equations into the flat six-dimensional
embedding space. One finds that the walls correspond to planes
\begin{eqnarray}
   & \displaystyle 
   \pm T_2 + k_1 \left(\frac{\rho_1}{H_1}\mp1\right) (T_1+Y) 
       = \frac{\rho_1}{\sqrt{-\lambda}H_1}
       & \qquad \textrm{for $y=0$}; \nonumber \\
   & \displaystyle 
   \pm T_2 + k_2 \left(\frac{|\rho_2|}{H_2}\mp1\right) (T_1+Y) 
       = \frac{|\rho_2|}{\sqrt{-\lambda}H_2}
       & \qquad \textrm{for $y=R$}. 
\label{eq:planes}
\end{eqnarray}
The choice of signs correspond to the choice of sign in the
expressions~(\ref{eq:gammai}) for $\gamma_1$ and $\gamma_2$. 

Note that both these equations have the form, for $i=1,2$,
\begin{equation}
   n_i \cdot X = c_i
\label{eq:plane}
\end{equation}
where $n_i$ is a time-like unit vector and $c_i>1/\sqrt{-\lambda}$. 
Intersecting with the AdS$_5$ hyperboloid~(\ref{eq:adshyper}), one
finds that the domain walls are dS$_4$ submanifolds of the embedding
space of curvature 
\begin{equation}
   \label{eq:Ccurv}
   \left(c_i^2+\frac{1}{\lambda^2}\right)^{-1/2} = H_i,
\end{equation}
where we have substituted the particular form of $c_i$ from
equation~(\ref{eq:planes}). (This can be seen explicitly by using the
$SO(4,2)$ symmetry of  the embedding flat space to put
$\vec{n}=(1,0,\dots,0)$ and substituting in
eqn. (\ref{eq:adshyper})). Thus we see again how the domain walls are
indeed dS$_4$ surfaces in the AdS$_5$ space with curvatures $H_i$. 

To get a sense of the global structure of the solution consider the case
where $k_1=k_2=0$ and choose the upper signs in the
solution~(\ref{eq:planes}). The planes are then simply given by
$T_2=|\rho_i|/(H_i\sqrt{-\lambda})$. The intersection is sketched in
figure \ref{fig:ads}. Note that in this case, the branes never
intersect. The figure allows to understand how the curves
describing the domain walls would turn into the hyperbolas shown in
figure \ref{fig:hyper1b} if we let $\lambda\to 0$. Intuitively, we can
generate the solutions described above by taking the hyperbolas of the
previous section and ``pasting'' them on the hyperboloid as shown in
figure \ref{fig:ads}. We can use this intuition to realize that the
various parameters (e.g., $k$, $p$, etc.) describing the solution play
a similar role as in the case $\lambda=0$. However, we should also
note that, in general, there is a second class of solutions where we
make the opposite choice of signs for the two walls
in~(\ref{eq:planes}). (Actually this choice is not possible for the
specific case $k_1=k_2=0$.) Now the planes are on ``opposite sides''
of the AdS hyperboloid. This has no natural $\lambda\to 0$ limit. It
corresponds to the result of section~\ref{dynamics} that we were
forced to choose a correlated signs in $\gamma_i$ to get the
flat-space solution.  

As a second example, we can briefly mention that the domain walls in
the Randall--Sundrum scenario (taking the $|\rho_i|\to\sqrt{-\lambda}$
limit in eqns. (\ref{eq:planes})) are described by the null planes  
\begin{eqnarray}
   & \displaystyle
   T_1 + Y = \frac{1}{k_1\sqrt{-\lambda}} 
      & \quad \textrm{for $y=0$}, \nonumber \\
   & \displaystyle
   T_1 + Y = \frac{1}{k_2\sqrt{-\lambda}} 
      & \quad \textrm{for $y=R$}, 
\end{eqnarray}
Again these branes will never intersect. 

To cast the above discussion in a more general set-up, recall that in
the previous section, the relative position of domain walls was
characterized in flat space by their Minkowskian distance (more
formally, by the distance between their asymptotic light
cones). Furthermore, in general this separation was null for the
solutions of section~\ref{soln}. Similarly here, the asymptote of each
dS in the six-dimensional embedding space is a light cone, whose
origin lies at  
\begin{equation}
   a_i = c_i n_i \qquad \textrm{for $i=1,2$}.
\end{equation}
Note that, in general, the vector $a_i$ will not lie on the bulk
AdS hyperboloid. We can now use the distance between the light cones,
given by  
\begin{equation}
   D\equiv a_1-a_2, 
\label{eq:sep}
\end{equation}
to characterize the relative location of the dS
surfaces in a coordinate-independent way. Furthermore, it is easily
seen that the quantity $D$ does reduce to the distance between
hyperboloids in the limit $\lambda\rightarrow 0$, as defined in
section \ref{dynamics}. Just as in the case $\lambda=0$ where our
solutions all described null separated hyperboloids, similarly, it can
be easily verified from equations~(\ref{eq:planes}) that all our
$\lambda<0$ solutions with a flat limit yield $D^2=0$, corresponding
to null separation. The size of the separation is controlled by the
parameters $k_1$ and $k_2$. 

Once again, we expect that solutions with time-like and
space-like separated dS surfaces are also allowed. We shall discuss
these more general configurations in section \ref{gen}.  

Finally, we note that the dynamics of the orbifold as well as the
causal properties of the solution are qualitatively the same as for
the case $\lambda=0$. The nature of the intersection between the walls
is controlled by $n_i$ and $\rho_i$. For general $n_i$, the
intersection is a null paraboloid. As before there are two cases, one
where the space-time has either expands from an initial singularity and
one which contracts to a final singularity. For the special case where
the $n_i$ are parallel or anti-parallel, there is no intersection and
the distance between the walls is fixed. For example, in the case
$-\rho_2>\rho_1>0$, we find that the orbifold remains static if the
initial distance between the branes equals $R_{static}$ (see
eqn. (\ref{eq:statickal})) and expands (collapses) if it is larger
(smaller) than $R_{static}$. Furthermore, two-way communication is
only possible for a finite amount of time if the extra dimension is
expanding.

\section{General embedded solutions} \label{gen}

In this section we will step back a little and reconsider the
solutions given thus far. In doing so we will see that they are
in fact special cases of a more general configuration of a pair of
branes in bulk AdS space. Note that in this section our solutions will
also no longer be confined to a particular coordinate patch of AdS
space.  

From the discussion of previous section, the solutions we found
correspond to a pair of dS surfaces embedded in AdS space. The dS
surfaces are not in general position but are null separated in the
sense discussed below eqn.~(\ref{eq:sep}). Recall that the solutions were
found by solving the bulk Einstein equation with negative cosmological
constant together with the Israel matching conditions
describing the discontinuity in the normal derivative of the metric at
the brane. 

The important point to note is that once we fix the bulk solution, in
this case to be AdS, the solution of the Israel conditions for the two
walls are completely independent. Each set of conditions are local
equations relating the shape of the brane embedded in the bulk space
to the stress energy on that brane, independent of the second
brane. From the analysis of the previous two sections, it appears that
for pure cosmological constant $\rho$ on the brane, the solution to
the Israel conditions is that the brane describe a dS surface embedded
in AdS. If this is correct, it is then clear that the general solution
corresponds to a pair of dS brane embedded in AdS with arbitrary
separation.  

To see that this is indeed the case, we can consider the Israel conditions
in a coordinate independent form and show that, when the bulk space is 
AdS$_5$, they imply that the brane is a dS$_4$ surface. If we let 
$t^{\mu}$ be the normal vector to the brane, the induced metric
$g^4_{\mu\nu}$ is then given by 
\begin{equation}
   g_{\mu\nu}=g^4_{\mu\nu}+t_{\mu}t_{\nu},
\end{equation}
where $g_{\mu\nu}$ is the bulk metric. As has been noted in various
papers (see e.g., \cite{chamblin}), assuming $Z_2$ reflection
invariance at the brane, 
the Israel conditions relate the extrinsic curvature $K_{\mu\nu}$ of
the brane to the stress energy $T^B_{\mu\nu}=-6\rho g^4_{\mu\nu}$ on
the brane. One has  
\begin{equation}
   K_{\mu\nu} = -\frac{1}{2}\left(
          T^B_{\mu\nu} - \frac{1}{3}T^B g^4_{\mu\nu} \right)
      = -\rho g^4_{\mu\nu}, 
\label{Israel}
\end{equation}
where the extrinsic curvature is given by
\begin{equation}
   K_{\mu\nu} = g_{\ \mu}^{4\ \rho}g_{\ \nu}^{4\ \kappa}
      \nabla_{\rho}t_{\kappa}.
\end{equation}
Since $g^4_{\mu\nu}$ and $t_\mu$ are functions of the embedding,
equation~(\ref{Israel}) is a local differential equation for the
functions describing the embedding of the brane in AdS$_5$, completely
independent of the presence of a second brane.  

To show that a dS$_4$ surface satisfies the Israel conditions, we
could evaluate~(\ref{Israel}) in a particular set of
coordinates. This is essentially what was done in section
2. Alternatively, we can argue simply by
symmetry that if the brane is dS the extrinsic curvature must be
proportional $g^4_{\mu\nu}$, since this is the only symmetric tensor
on the brane with the correct symmetries. The only question is then
what curvature of the dS space must we choose to make the constant of
proportionality exactly that in~(\ref{Israel}). To answer this we
recall that there is an expression for the intrinsic curvature $R_4$
of $g^4_{\mu\nu}$ in terms of the bulk curvature $R$ and
$K_{\mu\nu}$. In particular, we have the general expression
\begin{equation}
   R^4_{\kappa\lambda\mu\nu} = 
       g_{\ \kappa}^{4\ \kappa'}g_{\ \lambda}^{4\ \lambda'}
          g_{\ \mu}^{4\ \mu'}g_{\ \nu}^{4\ \nu'}R_{\kappa'\lambda'\mu'\nu'}
       + K_{\kappa\mu}K_{\lambda\nu} 
       - K_{\kappa\nu}K_{\lambda\mu}.
\end{equation}
Substituting the form of $K_{\mu\nu}$ and the bulk AdS space curvature
$R_{\kappa\lambda\mu\nu}=\lambda\left(
g_{\kappa\mu}g_{\lambda\nu}-g_{\kappa\nu}g_{\lambda\mu}\right)$ gives
the intrinsic scalar curvature 
\begin{equation}
   R_4 = 12\left(\rho^2+\lambda\right).
\label{R4}
\end{equation}
We have reproduced the result we derived in section 2. The curvature
of the brane dS space is such that the square of the Hubble
constant (see eq. (\ref{eq:binetruy})) is $\rho^2+\lambda$. 

\begin{figure}[ht]
\begin{center}
   \epsfig{file=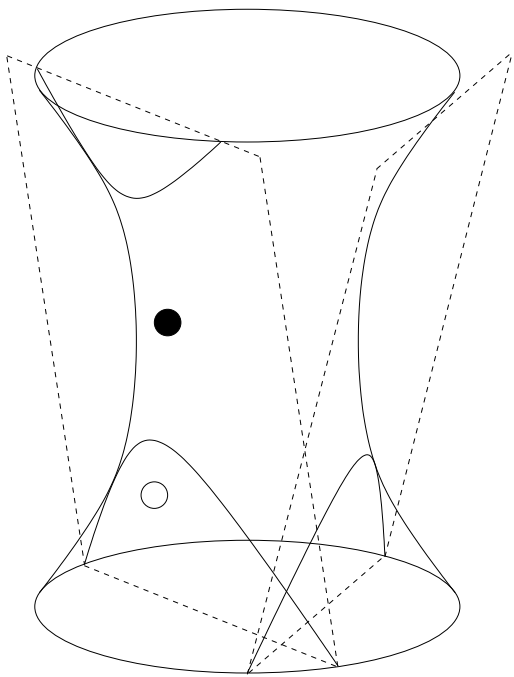,height=3.5in}
   \caption{General dS$_4$ branes in AdS$_5$ space formed by the intersection of planes 
with the AdS hyperboloid. Consider the brane formed by the left-hand plane. 
The sign of its tension, $\rho$, determines where the bulk lies. If $\rho<0$, 
the bulk includes the throat region, so that the solid circle lies in the bulk. 
If $\rho>0$, the bulk excludes the throat and, for example, the open circle lies 
in the bulk.}
   \label{fig:hyper}
\end{center}
\end{figure}

One notes that the curvature~(\ref{R4}) of the brane is independent of
the sign of the brane tension $\rho$. What then distinguishes the
$\rho>0$ case from $\rho<0$? Since the brane is a codimension-one
boundary in AdS space, we can either take the bulk space-time to be the
space ``inside'' the dS boundary or ``outside'' the dS
boundary. Consider Figure~\ref{fig:hyper}, which shows the
intersection of two planes with the AdS hyperboloid. Let us focus on
the left-hand plane, corresponding to a single dS submanifold. By ``inside''
we mean that the bulk space-time includes the throat of the AdS
hyperboloid. By ``outside'' we mean that the bulk space-time is one of the
two disconnected regions which do not include the throat. For example, the solid circle in figure \ref{fig:hyper} lies "inside" the dS boundary, while the open circle lies "outside" it. These two regions are
distinguished by the direction of the normal vector
$t_\mu$. Furthermore they have opposite extrinsic curvature
$K_{\mu\nu}$. It is then easy to show that one has the following
conditions  
\begin{equation}
\begin{array}{c}
   \textrm{if $\rho>0$ then the spacetime is ``inside'' the dS boundary} \\
   \textrm{if $\rho<0$ then the spacetime is ``outside'' the dS boundary} .
\end{array}
\end{equation}

We can now give a geometrical description of the general embedding
solution. As we argued above, since the Israel conditions are local we
can choose the branes to lie on \textit{any} pair of dS surfaces. As
noted in the previous section, in general, these are described by the
intersection of an arbitrary pair of planes with the AdS
hyperboloid. This is shown in Figure~\ref{fig:hyper}. For generic
choices of $n_1$ and $n_2$ the dS spaces will always intersect
transversally. This means there are configurations for {\it all values} of
the signs of $\rho_1$ and $\rho_2$. However, if  $n_1$ and $n_2$ are
either parallel or anti-parallel this is no longer true. In these cases
the branes never intersect. As a result, if the vectors are parallel,
one only has a solution with a bulk space-time bounded by a pair of
branes if $-\rho_1>\rho_2>0$ or $-\rho_2>\rho_1>0$. In the case where
they are anti-parallel one requires $\rho_1>0$ and $\rho_2>0$.  

The analysis of sections 3 and 4 allowed us to realize that the solutions
obtained in section 2 all described de Sitter surfaces separated by a null
vector. The above discussion has shown that time-like as well as
space-like separation vectors are also allowed. Furthermore, given that
two time-like vectors of equal magnitude are related by a boost, the
degrees of freedom describing the general solution are: the brane
tensions, the magnitude of the separation vector, and whether it is null,
space-like, or time-like. In addition, we note that, except for the
special case where the planes describing the dS branes are parallel or
anti-parallel, there are in general no conditions on the signs of $\rho_1$ and
$\rho_2$ for a solution to exist. The condition we found in section 2
is an artifact of using a particular coordinate system. 

The use of a specific coordinate system to analyze the null-separated
solutions was useful to describe the dynamics of the extra dimension as
viewed by an observer living on either brane (sec. 3.1) as well as the
appearance of horizons in the bulk (sec. 3.2). However, we noted that
these questions could also be addressed purely geometrically. The same
is true for the general embeddings described here. Of course, if
one wished, it is possible to describe the general solutions using
some global coordinate system adapted to one of the 
branes. As mentioned in section 2, one way to proceed would be to
start with the same ansatz for the metric as equation
(\ref{eq:ansatz}) but allow for a more general location of the second
brane.

We end this section by noting that although discussed in the context
of negative bulk cosmological constant $\lambda$ and dS branes
($\rho_i$ satisfying $\rho_i^2+\lambda>0$), the derivation of the
embedding conditions is completely general. The construction naturally
goes over to cases of positive or zero $\lambda$ and arbitrary
$\rho_i$. For instance, in bulk dS or flat space the intrinsic brane
curvature~(\ref{R4}) is always positive and we are considering embedded
dS branes. In AdS space, the intrinsic curvature can also be zero (the
Randall--Sundrum case) or negative. In the latter case we are
embedding AdS branes in the AdS bulk. Geometrically these arise from
intersecting with planes where $n$ is null (for flat branes) or
space-like (for AdS branes).

\section{Discussion} \label{discu}

All the solutions presented in this work were obtained by assuming
(albeit implicitly) that the bulk was AdS. It was then found that a
domain wall of uniform energy density can be embedded in this
background provided it follows a de Sitter trajectory, and we
described the most general configuration with two such trajectories in
AdS. Rather than fixing the bulk geometry for all times, a more
general approach is to treat the problem as an initial-value
problem. Let us assume that the only stress-energy in the problem is a
negative cosmological constant $\lambda$ in the bulk and brane
tensions $\rho_i$ such that $|\rho_i|>\sqrt{-\lambda}$. Suppose we
choose some space-like hypersurface on which we specify the initial
spatial bulk metric and the boundary branes. In general, one could
imagine complicated initial conditions, but a natural configuration to
consider is one with a space-like slice of anti-de Sitter spatial bulk
metric and two (spatially) flat surfaces with arbitrary separation,
velocity, and orientation. If one were to solve for the time evolution
of this system, one would generically find that the bulk does not
remain AdS but that a non-vanishing Weyl tensor is generated. This
would mean that our solutions correspond to initial conditions which
are tuned so that the bulk remains AdS. Coming back to the general
problem, it is not clear whether all bulk evolutions would yield
homogeneous and isotropic cosmologies on the branes. It would be
essential to know what subclass of initial conditions would be
consistent with the usual assumptions of cosmology. 

A related issue has to do with the stability of the solutions. Suppose
our tuned initial conditions are perturbed, then one should
investigate whether the path is only slightly or greatly disturbed by
the variation. The brane motion could be perturbed either by moving
the brane as a whole away from its initial trajectory without altering
its energy density, by moving a region of the brane off the trajectory
while keeping the energy density constant, or by perturbing the energy
density in a spatially homogeneous or inhomogeneous way. In the first
case, we know of at least one example of instability, namely solutions
with static orbifolds in the case $\rho_1\neq -\rho_2$. Indeed, we saw
in section \ref{dynamics} that the branes will eventually collide if
brought infinitesimally closer than the static distance,  or end up
infinitely far apart if pulled away from each other. Note that, even
though the path is unstable, the cosmological evolution remains de
Sitter and the bulk is still AdS. Nevertheless, more general homogeneous 
perturbations may drive the bulk away from AdS and, thus, 
drive the cosmological evolution on each brane away from dS. As for 
inhomogeneous energy density perturbations, it was argued in Ref. \cite{shiro} that any
spatial inhomogeneities on the brane stress-energy will modify the AdS
bulk through gravitational radiation. Cosmological energy density
perturbations in brane-worlds have recently been investigated in
Refs. \cite{muko, koda, langlois, bruck}. 

To conclude, let us briefly point out that our solutions can be
straightforwardly generalized to any type of energy density on the
brane (e.g., radiation, matter). Fixing the bulk metric to be AdS for
simplicity, one can use the Israel junction condition at the location
of the brane to solve for its motion in the bulk
\cite{chamblin,krauss}, which in turn determines its cosmological
evolution. If we want to embed two domain walls, each one should be
allowed to travel along any trajectory consistent with its energy
density.

\bigskip
\noindent 
{\sc Acknowledgments} \\
We would like to thank Burt Ovrut, Alexander Polyakov, Andy
Strominger, and Herman Verlinde for helpful discussions. J.K. is
supported by the Natural Sciences and Engineering Research Council of
Canada. P.J.S. is supported in part by US Department of Energy grant
DE-FG02-91ER40671. D.W. would like to thank The Rockefeller University and
The University of Chicago for hospitality during the completion of
this work. 

\pagebreak

\pagebreak

\end{document}